\documentclass[preprint,3p]{elsarticle}

\makeatletter
\def\ps@pprintTitle{%
 \let\@oddhead\@empty
 \let\@evenhead\@empty
 \def\@oddfoot{\centerline{\thepage}}%
 \let\@evenfoot\@oddfoot}
\makeatother

\usepackage{amsmath,amssymb}
\usepackage[T1]{fontenc}
\usepackage{url}
\usepackage{tabularx}
\usepackage{varwidth}
\usepackage{wrapfig}
\usepackage{graphics}
\usepackage{graphicx}
\usepackage[usenames, dvipsnames]{xcolor}
\usepackage{color, colortbl} 
\usepackage{adjustbox}
\usepackage{stackengine}
\usepackage{multirow}
\usepackage{lineno}
\usepackage{booktabs}
\usepackage{tabu}
\usepackage{makecell}
\usepackage{relsize}
\usepackage[colorlinks=true,linkcolor=black,filecolor=blue,citecolor=red,bookmarksnumbered=true]{hyperref}
\usepackage{filecontents}
\usepackage{enumitem}
\usepackage{caption}
\usepackage{subcaption}

\usepackage{setspace}

\journal{International Journal of Multiphase Flow}

\newcommand{\ub}{\mathbf{u}}

\newcommand{\M}{\mathbf{M}}
\newcommand{\F}{\mathbf{F}}
\newcommand{\x}{\mathbf{x}}
\newcommand{\at}{\textsf{@}}

\begin{document}


\begin{frontmatter}

\title{Quantifying the errors of the particle-source-in-cell Euler-Lagrange method\vspace{-5mm}}%

\author{Fabien Evrard}\corref{cor1}
\cortext[cor1]{Corresponding author}
\ead{fabien.evrard@ovgu.de}
\author{Fabian Denner}
\author{Berend van Wachem}

\address{Lehrstuhl f\"ur Mechanische Verfahrenstechnik, Otto-von-Guericke-Universit\"at Magdeburg, \mbox{Universit\"atsplatz 2, 39106 Magdeburg, Germany}}

\begin{abstract}
The particle-source-in-cell Euler-Lagrange (PSIC-EL) method is widely used to simulate flows laden with particles. Its accuracy, however, is known to deteriorate as the ratio between the particle diameter~($\smash{d_\text{p}}$) and the mesh spacing~($h$) increases, due to the impact of the momentum that is fed back to the flow by the Lagrangian particles. Although the community typically recommends particle diameters to be at least an order of magnitude smaller than the mesh spacing, the errors corresponding to a given $\smash{d_\text{p} / h}$ ratio and/or flow regime have not been systematically studied. In~this paper, we provide an expression to estimate the magnitude of the flow velocity disturbance resulting from the transport of a particle in the PSIC-EL framework, based on the $\smash{d_\text{p} / h}$ ratio and the particle Reynolds number, $\smash{\text{Re}_\text{p}}$. This, in turn, directly relates to the error in the estimation of the undisturbed velocity, and therefore to the error in the prediction of the particle motion. We show that the upper bound of the relative error in the estimation of the undisturbed velocity, for all particle Reynolds numbers, is approximated by $\smash{(6/5)\,d_\text{p} / h}$. Moreover, for all cases where $\smash{d_\text{p} / h \lesssim 1/2}$, the expression we provide accurately estimates the value of the errors across a range of particle Reynolds numbers that are relevant to most gas-solid flow applications ($\smash{\text{Re}_\text{p} < 500}$). \\~\\
{\small \textcopyright~2020. This manuscript version is made available under the CC-BY-NC-ND 4.0 license.\\
\href{http://creativecommons.org/licenses/by-nc-nd/4.0/}{http://creativecommons.org/licenses/by-nc-nd/4.0/}\\
This manuscript has been accepted for publication in the {\em International Journal of Multiphase Flow}.\\
\href{https://doi.org/10.1016/j.ijmultiphaseflow.2020.103535}{https://doi.org/10.1016/j.ijmultiphaseflow.2020.103535}\\
Please refer to the journal version when citing this work.}
\end{abstract}
\end{frontmatter}

\section{Introduction}
The Euler-Lagrange (EL) method has been used to simulate the behavior of particle-laden flows for many years \cite{Eaton2009,Balachandar2009,Balachandar2010,Kuerten2016}. While tracking particles individually, the EL method does not resolve the no-slip condition at the surface of each particle. This means that the fluid forces acting on each particle have to be calculated based on the local underlying flow, using reduced models. For particle volume fractions as low as $10^{-5}$, the transfer of momentum from the particles to the fluid is significant enough to alter the flow, and therefore needs to be considered \cite{Elghobashi1994}. Approaches that do so are referred to as \textit{two-way coupled}: the motion of the particles is dictated by the underlying flow field, which itself is impacted by the momentum that is fed back by the particles -- hence a two-way coupling between the fluid and particulate phases.
\subsection{Governing equations} When modelling dilute particle-laden flows within a two-way coupled EL framework, the equations governing the fluid phase are the Navier-Stokes equations 
\begin{align}
		\nabla \cdot \ub_{\text{f}}  & = 0  \, , \label{eq:nsvectorcont} \\
	  \rho_{\text{f}} \left[\dfrac{\partial \ub_{\text{f}}}{\partial t} + \nabla \cdot (\ub_{\text{f}} \otimes \ub_{\text{f}}) \right] &  = -\nabla p + \mu_{\text{f}} \, \Delta \ub_{\text{f}} + \M \, , \label{eq:nsvectormom} 
\end{align}
where $\ub_{\text{f}}$ is the fluid velocity vector, $\rho_\text{f}$ the fluid density, $\mu_\text{f}$ the fluid viscosity, $p$ the pressure, and to which $\M$, a term corresponding to the transfer of momentum from the particles to the fluid, is added.
The motion of a spherical particle is described by Newton's second law
\begin{equation}
	\rho_\text{p} \dfrac{\pi\,d_\text{p}^3}{6} \dfrac{\text{d}\ub_\text{p}}{\text{d}t} = \F_\text{p,fluid} + \F_\text{p,body} \, ,
\end{equation}
where $\rho_\text{p}$ is the particle density, $d_\text{p}$ the particle diameter, $\ub_{\text{p}}$ the particle velocity vector, $\F_\text{p,fluid}$ the resultant of the fluid forces acting on the particle surface, and $\F_\text{p,body}$ the resultant of the body forces acting on the particle. The resultant of the fluid forces acting on the particle surface can be expressed as a function of the local \textit{undisturbed} flow based on the Maxey-Riley-Gatignol equation \cite{Maxey1983,Gatignol1983}. It then reads as
\begin{equation}
    \F_\text{p,fluid} = \F_\text{p,drag} + \F_\text{p,undist} + \F_\text{p,add} + \F_\text{p,hist} \, ,
\end{equation}
where $\F_\text{p,drag}$ is the quasi-steady drag force, $\F_\text{p,undist}$ the undisturbed flow force, $\F_\text{p,add}$ the added-mass force, and $\F_\text{p,hist}$ the Basset history force. When the particles are significantly denser than the fluid ($\rho_\text{p} \gg \rho_\text{f}$), drag is the main fluid force contribution, and other force contributions can be neglected ($\F_\text{p,fluid} \simeq \F_\text{p,drag}$)~\cite{Boivin1998,Armenio2001}. The equation for the drag acting on a spherical particle reads as
\begin{equation}
    \F_\text{p,drag} = 3\,\pi\,\mu_\text{f}\,d_\text{p}\,f\!\left(\text{Re}_\text{p}\right)\left(\tilde{\ub}_\text{f\:\!\at\:\!\text{p}} - \ub_\text{p}\right) \, ,
\end{equation}
which corresponds to Stokes' law augmented with an empirical factor $\smash{f\!\left(\text{Re}_\text{p}\right)}$ in order to consider finite values of the particle Reynolds number, $\smash{\text{Re}_\text{p}}$. The drag force depends on $\smash{\tilde{\ub}_\text{f\:\!\at\:\!\text{p}}}$, the \textit{undisturbed} fluid velocity associated with the particle. This is defined as the local fluid velocity as though the particle had been taken out of the flow. The estimation of the undisturbed velocity is notoriously difficult, as it involves having access to an equivalent, conceptual flow where the particle under consideration does not exist \cite{Boivin1998,Balachandar2009,Balachandar2019}. In a two-way coupled EL framework, specifically, the fluid velocity at the location of each particle features a disturbance that is induced by the momentum transferred from the particle to the fluid. Determining the undisturbed velocity would thus formally require to solve the governing flow equations without transfer of momentum from the particle under consideration, as many times as there are particles in the flow. Since this is, in practice, impossible, the undisturbed velocity is almost always approximated by the \textit{disturbed} or \textit{actual} fluid velocity interpolated to the position of the particle. The community has only very recently started to propose models for the estimation of the undisturbed velocity \cite{Gualtieri2015,Horwitz2016,Ireland2017,Horwitz2018,Balachandar2019,Evrard2020b}, mostly derived within the framework of the \textit{volume-filtered} of \textit{regularised} EL method, but to date, their use remains marginal.

\subsection{Numerics}
Assumptions are required to calculate $\M$, the transfer of momentum from the particles to the fluid. The particle-source-in cell (PSIC) model of \citet{Crowe1977}, although proposed in the 1970's, is still widely used to address this issue, owing to its relative simplicity compared to the more recent approaches that consist in convoluting the momentum contribution with a smooth kernel \cite{Capecelatro2013,Ireland2017,Balachandar2019,Poustis2019,Evrard2020b}. In a computational cell $K$, the momentum transfer term as defined in the PSIC model reads as
\begin{equation}
	\M_K 
    = \sum_{\text{p} \, \in \, K} -\left(\F_\text{p,drag}+\F_\text{p,add}+\F_\text{p,hist}\right)  \, \delta_K(\x_\text{p}) \, ,
\end{equation}
where $\smash{\x_\text{p}}$ is the centre of the particle $\text{p}$ and $\smash{\delta_K}$ is a piecewise-constant approximation of the Dirac delta function whose compact support is the computational cell $K$ where the particle is located. Assuming that the cell has a volume $\smash{V_K}$, then $\smash{\delta_K}$ is defined as
\begin{equation}
	\delta_K(\x) = \left\{ \begin{array}{ll} 1 / V_K & \text{$\x$ $\in$ cell} \\ 0 & \text{elsewhere}\end{array}\right. \, .
\end{equation}
It is known that the PSIC-EL framework is not numerically convergent \cite{Garg2007,Garg2009,Poustis2019}, and that the errors due to a particle's self-induced flow disturbance grow proportionally with the ratio $\smash{\hat{d}_\text{p}}$ between the particle diameter $\smash{d_\text{p}}$ and the mesh spacing $\smash{h = V_K^{1/3}}$ \cite{Boivin1998,Gualtieri2013}. To mitigate these errors, \textit{i.e.} keep the magnitude of the local velocity disturbance small, a stringent constraint is usually imposed: the tracked Lagrangian particles have to be ``much'' smaller than the computational cells in which the governing flow equations are solved. The consensus in the community seems to be $\smash{\hat{d}_\text{p} \lesssim 0.1}$. Yet, in the framework of the PSIC-EL method, there is no systematic study quantifying the errors associated with such ratios, and/or on the impact of the flow regime (\textit{e.g.} the impact of $\smash{\text{Re}_\text{p}}$, the Reynolds number of the tracked particle) on these errors.

\subsection{Outlook}
In this paper, we present an expression to estimate the order of magnitude of the self-induced flow velocity disturbance of a particle tracked in the PSIC-EL framework, based on the ratio $\smash{\hat{d}_\text{p} = d_\text{p} / h}$ and the particle Reynolds number $\smash{\text{Re}_\text{p}}$. Its derivation relies on an approximate solution to the Navier-Stokes equations, using Oseen's approximation. Our aim is to provide the community with a way to assess the errors that are to be expected when simulating particle-laden flows within the two-way coupled PSIC-EL framework, for a range of $\smash{\hat{d}_\text{p} = d_\text{p}/h}$ and $\smash{\text{Re}_\text{p}}$ values that are relevant to the majority of gas-solid flow applications. The derived expression is then validated with the results of numerical simulations in order to assess its predictive abilities.
 
\section{Discrete flow disturbance due to a singular momentum contribution}
\label{sec:derivation}
\begin{figure}
    \adjincludegraphics[width=1.0\textwidth, trim={{.0\width} {.18\height} {.0\width} {.25\height}}, clip, rotate=0]{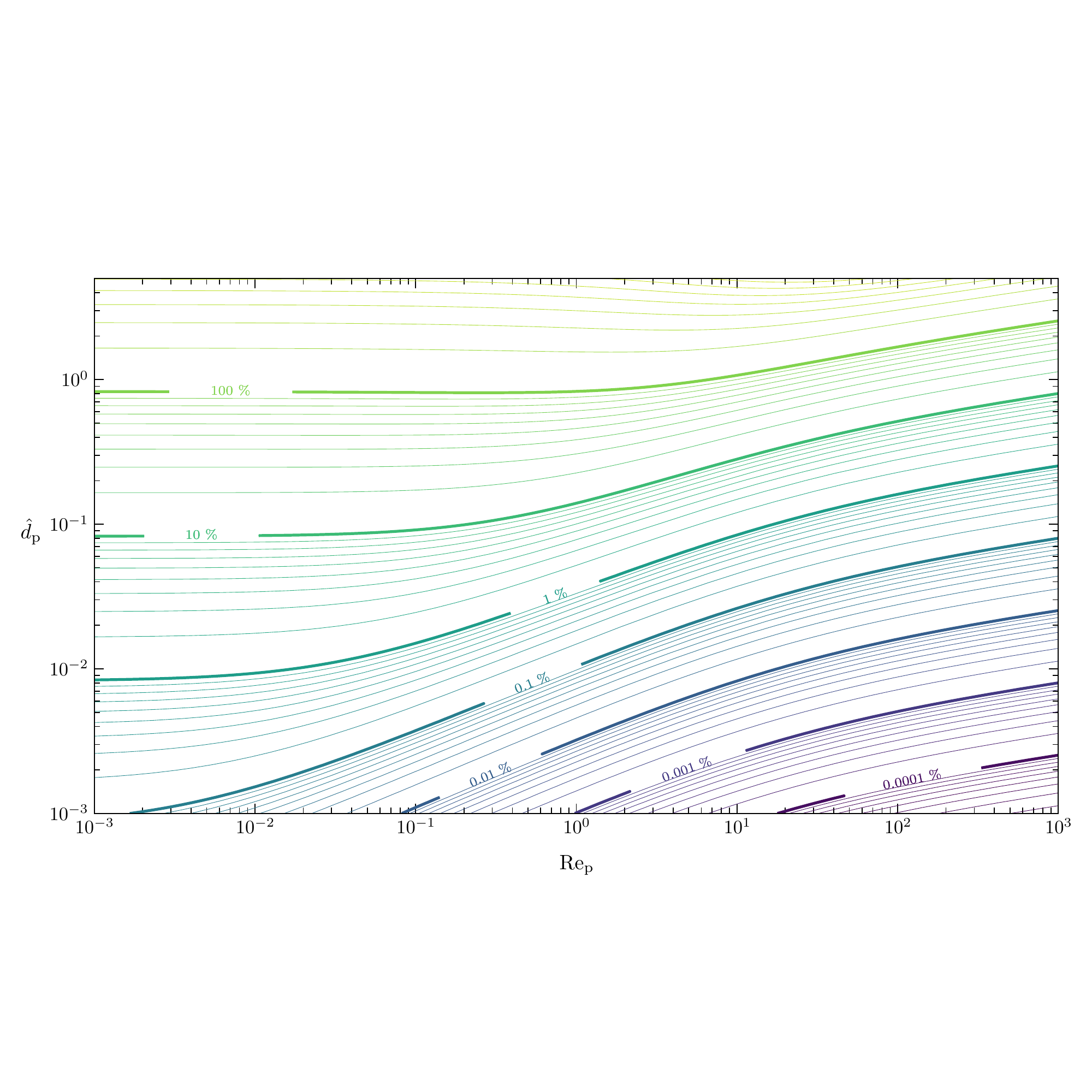}
    \caption{Contour map of $\smash{|u_{h} - u_\infty| / u_\infty}$, the estimated amplitude of the velocity disturbance normalised by the particle relative velocity, in the cell where the particle is located, as a function of the ratio between the particle diameter and the mesh spacing, $\smash{\hat{d}_\text{p} = d_\text{p} / h}$, and the particle Reynolds number, $\smash{\text{Re}_\text{p}}$. The finite Reynolds number drag correction function $\smash{f\!\left(\text{Re}_\text{p}\right)}$ employed is that of \citet{Schiller1933}.}
    \label{fig:map}
\end{figure}

Our aim is to estimate the magnitude of the local velocity disturbance associated with a particle modelled within the PSIC-EL framework, in order to estimate the error with which drag is computed. To that end, we consider the representative albeit simplified case of a single, isolated particle moving through quiescent flow at constant velocity. Without loss of generality, we can equivalently consider the flow with far field velocity $\textbf{u}_{\infty} = u_{\infty} \, \mathbf{e}_\text{x}$ against a single, fixed particle located at the centre of the domain and origin of the coordinate system -- this corresponds to the very same flow observed in the frame of reference of the particle. The resultant of the fluid forces acting on the particle is then also collinear to $\mathbf{e}_\text{x}$.
In this configuration, the undisturbed flow velocity is $\mathbf{u}_{\infty}$. Within the PSIC-EL framework, however, the flow velocity as seen by the particle differs from this undisturbed velocity $\mathbf{u}_{\infty}$, since the momentum that is fed back to the fluid via the transfer term $\M$ generates a local flow disturbance. This flow velocity, expressed in the frame of reference of the particle, is the solution to the Navier-Stokes equations~\eqref{eq:nsvectorcont} and \eqref{eq:nsvectormom} with the momentum transfer term\footnote{At this point in the derivation, we consider the solution to the flow in a continuous infinite domain, and therefore $\delta$ is the Dirac delta function; not its discrete counterpart $\delta_K$.} $\M = -F \, \delta(\mathbf{0}) \, \mathbf{e}_\text{x}$. Under the assumption that the magnitude of the velocity disturbance remains small compared to $u_\infty$, it is sound to linearise the convective term of the momentum equations~\eqref{eq:nsvectormom} with respect to the far-field velocity $\mathbf{u}_\infty$, in which case we consider what is commonly referred to as Oseen's approximation of the Navier-Stokes equations~\cite{Oseen1927}. Considering the steady nature of the case we study in this section, we can also neglect the transient term of the momentum equations~\eqref{eq:nsvectormom}. The $x$-component of the velocity field solution to these simplified equations is then given by~\cite{Venkatalaxmi2007}
\begin{equation}
    u = u_\infty - \dfrac{F}{8\,\pi\,\mu_\text{f}\,\rho_\text{f}\,u_\infty\,r^3} \left( 2\,\mu_\text{f}\,x - \left(u_\infty\,\rho_\text{f}\,r\,\left(x + r\right) + 2\,\mu_\text{f}\,x\right) \exp\left( \dfrac{\rho_\text{f}\,u_\infty(x-r)}{2\,\mu_\text{f}}\right) \right) \, ,
\end{equation}
where $x$ is the position on the $x$-axis, and $r$ the distance to the origin of the coordinate system (where the particle is located). The expression of the velocity disturbance $u - u_\infty$ is referred to as the \textit{Oseenlet}. Introducing the particle Reynolds number $\smash{\text{Re}_\text{p} = \rho_\text{f} \, u_\infty \, d_\text{p} / \mu_\text{f}}$, and considering that drag is the dominant fluid force acting on the particle, this then reads as
\begin{equation}
   u = u_\infty \left(1 + \dfrac{3\,d_\text{p}\,f\!\left(\text{Re}_\text{p}\right)}{8\,\text{Re}_\text{p}\,r^3} \left(  \left( \text{Re}_\text{p}\,r\,(r + x) +2\,d_\text{p}\,x \right)\exp\left( \dfrac{\text{Re}_\text{p} \left(x-r\right)}{2\,d_\text{p}} \right) -2\,d_\text{p}\,x \right) \right) \, .
   \label{eq:oseenlet}
\end{equation}
On a numerical mesh, in the framework of the PSIC model, the solution of this singular Oseen flow will produce a velocity $\smash{u_h}$ in the cell with volume $\smash{h^3}$ centred at the origin of the coordinate system (where the particle is located). This velocity $u_h$ can be expected to be an approximation of the analytical field defined in Eq.~\eqref{eq:oseenlet} averaged in a sphere of volume $\smash{h^3}$ centred at the origin\footnote{Although computational cells are classically hexahedral, or more generally polyhedral, we integrate the analytical flow over a sphere of similar volume as the cell, in order to take advantage of the axi-symmetric nature of the Oseenlet.}, that is
\begin{equation}
  u_h = \dfrac{2\,\pi}{h^3} \int_0^{\pi}\int_0^{\alpha\,h} u(r,\theta) \, r^2 \sin(\theta) \ \text{d}r \text{d}\theta \, ,
\end{equation}
with $\smash{\alpha = \left(3/(4\,\pi)\right)^{1/3}}$.
Note that, when expressed in spherical coordinates, $u$ does not depend on the azimutal coordinate since the flow is axi-symmetric about the $x$-axis. The solution to the previous integral reads as
\begin{equation}
  u_h = u_\infty \left(1 + \pi\,\alpha^2\,\hat{d}_\text{p}\,\Psi_\text{Oseen}\left( \text{Re}_\text{p}, \hat{d}_\text{p} \right)\right) \, ,
  \label{eq:disturbance}
\end{equation}
where $\smash{\hat{d}_\text{p} = d_\text{p} / h}$, and with the Oseen correction function
\begin{equation}
\Psi_\text{Oseen}\left( \text{Re}_\text{p}, \hat{d}_\text{p} \right) = 3\,\hat{d}_\text{p}\,f\!\left(\text{Re}_\text{p}\right) \left( (\alpha\,\text{Re}_\text{p})^{-1} 
- 2\, \hat{d}_\text{p}\,(\alpha\,\text{Re}_\text{p})^{-2} + 2\,\hat{d}_\text{p}^{\,2}\,(\alpha\,\text{Re}_\text{p})^{-3} \left(1 - \exp\left( -\dfrac{\text{Re}_\text{p}\,\alpha}{\hat{d}_\text{p}} \right) \right) \right) \, .
\label{eq:oseencorr}
\end{equation}
A contour map of the estimated relative amplitude of the velocity disturbance, $\smash{|u_{h} - u_\infty| / u_\infty}$, as a function of $\smash{\hat{d}_\text{p}}$ and $\smash{\text{Re}_\text{p}}$, is given in Figure~\ref{fig:map}. From the knowledge of $\smash{\hat{d}_\text{p}}$ and $\smash{\text{Re}_\text{p}}$ of a particle, this contour map allows to graphically estimate the error made when estimating the local undisturbed velocity, and therefore the error made when computing the drag force acting on the particle.
It can be shown that for all $\smash{\hat{d}_\text{p}}$, 
\begin{equation}
    \lim_{\text{Re}_\text{p} \to 0} \Psi_\text{Oseen}\left( \text{Re}_\text{p}, \hat{d}_\text{p} \right) = 1 \, ,
\end{equation}
therefore the local discrete velocity is approximated in the Stokes limit as
\begin{equation}
 u_h = u_\infty \left(1 + \pi\,\alpha^2\,\hat{d}_\text{p} \right) \simeq u_\infty \left(1 + (6/5)\,\hat{d}_\text{p} \right) \, .
\end{equation}In the Stokes limit, for a particle moving at constant velocity within the PSIC-EL framework, we should thus expect a local velocity disturbance whose amplitude is about $\smash{(6/5)\,\hat{d}_\text{p}}$ times that of the relative velocity of the particle. When $\smash{\hat{d}_\text{p} = 0.1}$ (which is often considered as the upper limit of validity of the PSIC-EL method), this corresponds to an error in the estimation of the local undisturbed velocity, using the local velocity as seen by the particle, of about $12\,\%$. An increase in the Reynolds number of the particle will generally result in a decrease of $\smash{\Psi_\text{Oseen}}$ (which converges towards zero when $\smash{\text{Re}_\text{p} \to \infty}$), meaning that in terms of the local velocity disturbance, in the range of validity of the Oseen approximation, the largest errors are found in the Stokes regime. 

\section{Validation with numerical results}
\label{sec:results}
The estimation of the discrete velocity disturbance proposed in Eq.~\eqref{eq:disturbance} is validated against numerical results. Simulations are conducted in our in-house finite-volume CFD research code \textit{MultiFlow} (second-order accurate in time and space) \cite{Denner2020}. The conducted test-cases consider the uniform flow through the momentum source field corresponding to the momentum transfer of a single particle located at the centre of the domain, using the assumptions of the PSIC model. The dimensions of the cubical flow domain are such that its width is $1000$ times larger than the width $h$ of the central computational cell in the domain. The particle Reynolds number is chosen randomly in the interval $[0.005,500]$, and the ratio $\smash{\hat{d}_\text{p} = d_\text{p}/h}$ is chosen randomly in $[0.005,5]$ (the distribution of these random numbers is uniform on the logarithmic scale). The finite Reynolds number drag correction function $\smash{f\!\left(\text{Re}_\text{p}\right)}$ employed is that of \citet{Schiller1933}. The domain is initialised with a flow of uniform velocity $\mathbf{u}_\infty$, and run until a steady state is reached (which is considered to happen when the relative difference in the normalised velocity disturbance is less than $10^{-6}$ from one timestep to the next). More than $N_\text{s} = 800$ test-cases with random $\smash{\hat{d}_\text{p}}$ and $\smash{\text{Re}_\text{p}}$ are conducted.\\

\begin{figure}
\centering
\begin{minipage}[t]{.475\textwidth}
  \centering
    \hspace{-0.025\textwidth}
    \includegraphics[height=1.3\textwidth]{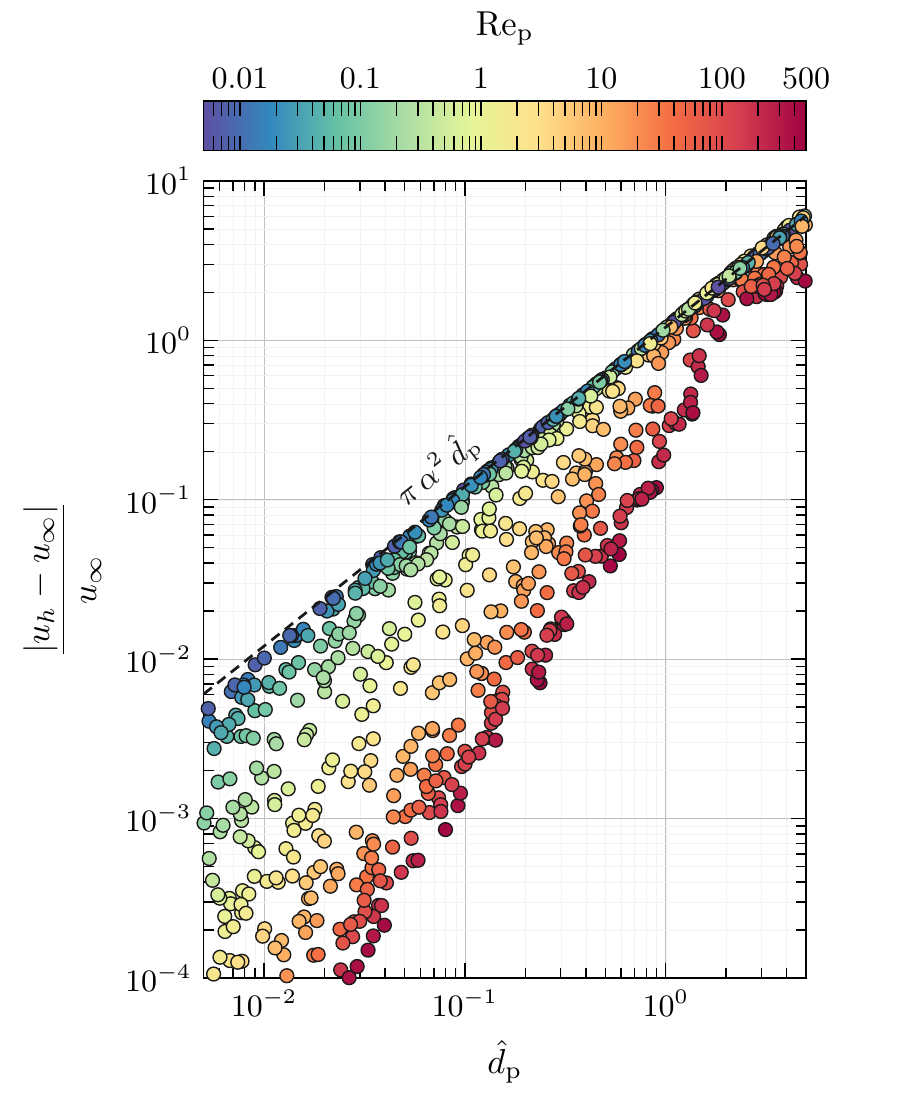}
\caption{Discrete velocity disturbances observed in the cell where the particle is located, using our numerical framework \textit{MultiFlow}, normalised by the far-field velocity $u_\infty$.}
\label{fig:errors}
\end{minipage}%
\begin{minipage}[t]{.05\textwidth}
~
\end{minipage}%
\begin{minipage}[t]{.475\textwidth}
  \centering
    \hspace{-0.025\textwidth}
    \includegraphics[height=1.3\textwidth]{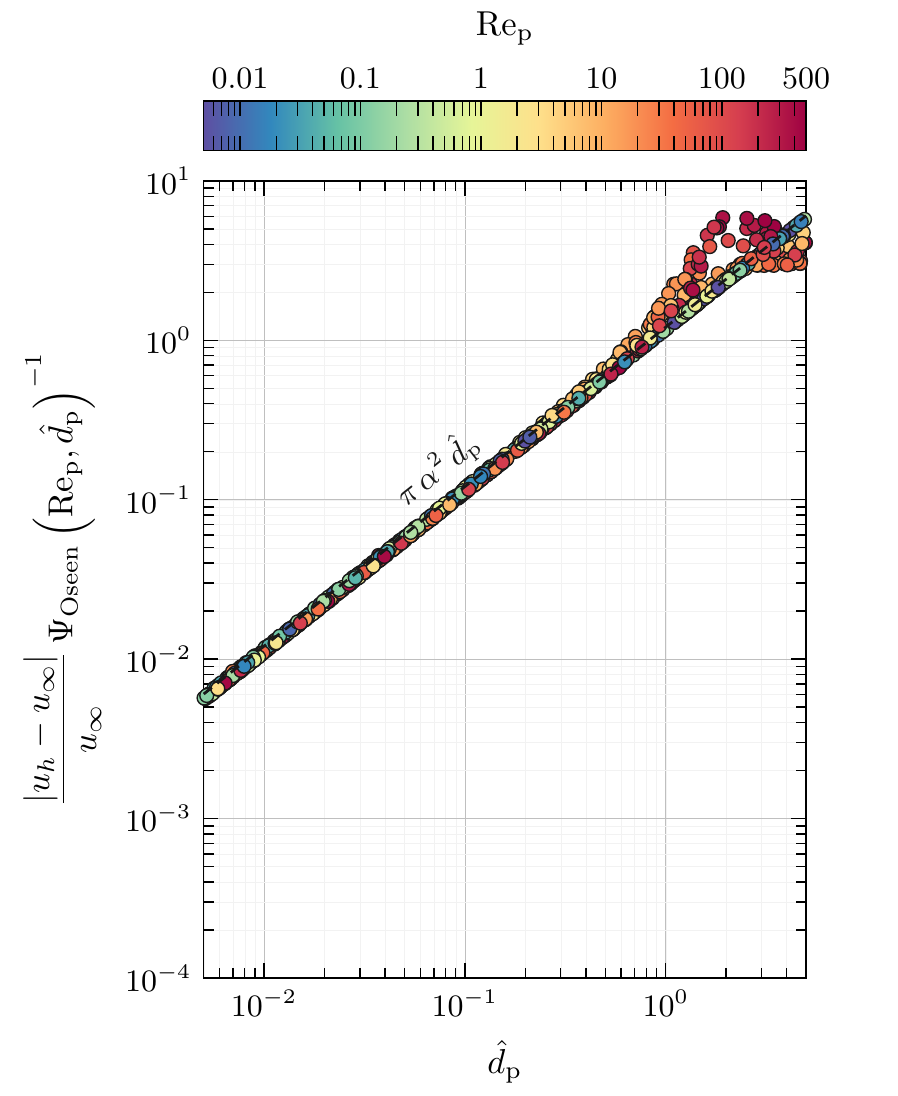}
\caption{Discrete velocity disturbances observed in the cell where the particle is located, using our numerical framework \textit{MultiFlow}, normalised by the far-field velocity $u_\infty$ multiplied by the Oseen correction function $\smash{\Psi_\text{Oseen}}$, Eq.~\eqref{eq:oseencorr}.}
\label{fig:errorsOS}
\end{minipage}
\end{figure}

Figures~\ref{fig:errors}~and~\ref{fig:errorsOS} shows the magnitude of the discrete velocity disturbance $\smash{|u_h - u_\infty|}$ normalised by the undisturbed velocity $u_\infty$, as well as the same data normalised by the Oseen correction function $\smash{\Psi_\text{Oseen}}$ defined in Eq.~\eqref{eq:oseencorr}, respectively. Figure~\ref{fig:errors} confirms that $\smash{\pi\,\alpha^2\,\hat{d}_\text{p} \simeq (6/5)\,\hat{d}_\text{p}}$ is a good approximation of the upper bound of the relative discrete velocity disturbance, for all Reynolds numbers considered. In Figure~\ref{fig:errorsOS}, where the relative error is normalised by the Oseen correction function $\smash{\Psi_\text{Oseen}}$, the proximity of the data points to the dashed line indicates the ability of our expression to account for the impact of the particle Reynolds number on the magnitude of the discrete velocity disturbance.
Deviation from the dashed line is observed for large values of $\smash{\hat{d}_\text{p}}$ and $\smash{\text{Re}_\text{p}}$. Such large values yield velocity disturbances that are similar in magnitude to the far-field velocity $u_\infty$: the validity of the Oseen approximation of the Navier-Stokes equations is therefore questionable for this region of the parameter map (as explained in Section~\ref{sec:discussionoseen}). Overall, for particles that are twice smaller than the mesh spacing (\textit{i.e.} $\smash{\hat{d}_\text{p} \lesssim 1/2}$), the proposed expression proves very capable of predicting the amplitude of the flow disturbance associated with a particle moving in the PSIC-EL framework, across the wide range of Reynolds numbers considered.

\section{On the limits of Oseen's approximation and alternative scaling}
\label{sec:discussion}
\subsection{Limits of Oseen's approximation}
\label{sec:discussionoseen}
    Oseen's approximation consists of the linearisation of the convective term of the Navier-Stokes equations with respect to the far-field velocity and should, in principle, only be valid for small values of the Reynolds number, \textit{i.e.} $\text{Re} \ll 1$. In Figure~\ref{fig:errorsOS}, we observe that for $\smash{\hat{d}_\text{p} \gtrsim 1/2}$ and $\smash{\text{Re}_\text{p} \gtrsim 1}$, the results of the numerical simulations do not collapse onto the dashed line, which seems to be consistent with the previous statement. For $\smash{\hat{d}_\text{p} \lesssim 1/2}$, however, all points collapse onto the dashed line, even for values of the particle Reynolds number as high as $\text{Re}_\text{p} = 500$. 
    This seemingly surprising result can in fact be expected if one considers the flow \textit{actually} solved on the Eulerian grid. In the framework of the PSIC-EL method, the Eulerian flow solver does not ``see'' the particle \textit{per se}, \textit{i.e.} it does not enforce a no-slip condition at the particle boundary; instead, it perceives the momentum source $\M$, whose regularisation support scales with the mesh-spacing $h$. The Reynolds number with regard to the width of the momentum source support, $h$, and the magnitude of the velocity disturbance, $\smash{\left|u_h-u_\infty\right|}$, is therefore the one that is relevant for discussing the validity of Oseen's approximation. It is given by
    \begin{equation}
        \text{Re}_{h} = \dfrac{\rho_\text{f}\,\left|u_h-u_\infty\right|\,h}{\mu_\text{f}} = \text{Re}_\text{p} \dfrac{\left|u_h-u_\infty\right|}{\hat{d}_\text{p} \, u_\infty} \, .
    \end{equation}
    When Oseen's approximation is valid, the following approximation, derived from Eq.\eqref{eq:disturbance}, holds:
    \begin{equation}
         \text{Re}_{h} \simeq \pi\,\alpha^2\,\Psi_\text{Oseen}\left( \text{Re}_\text{p}, \hat{d}_\text{p} \right)\text{Re}_\text{p} \, . \label{eq:reh}
    \end{equation}
    \begin{figure}
\centering
\begin{minipage}[t]{.475\textwidth}
  \centering
    \hspace{-0.025\textwidth}
    \includegraphics[height=1.3\textwidth]{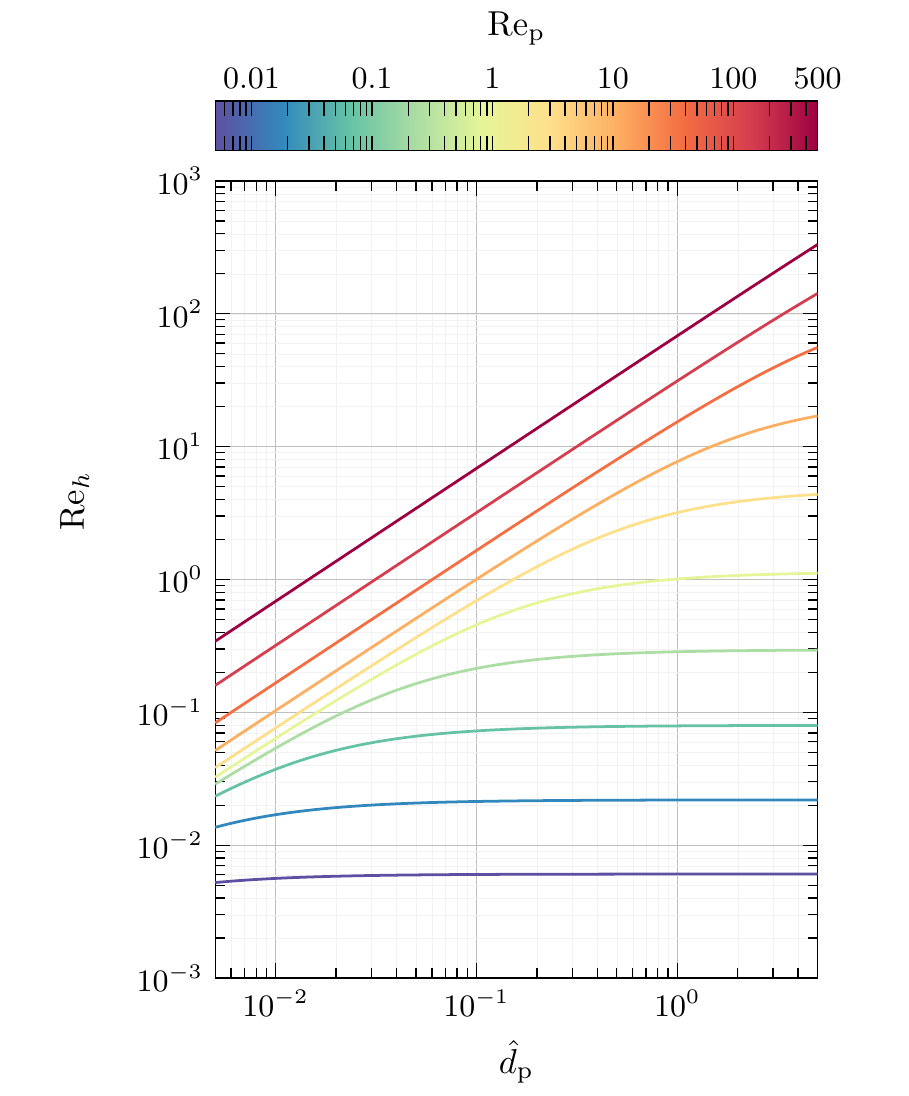}
    \captionof{figure}{Evolution of $\smash{\text{Re}_h}$, the Reynolds number with regard to the width of the momentum source support, $h$, and the magnitude of the velocity disturbance, $\smash{\left|u_h-u_\infty\right|}$, as a function of the particle Reynolds number $\smash{\text{Re}_\text{p}}$ and the the ratio between the particle diameter and the mesh spacing, $\smash{\hat{d}_\text{p} = d_\text{p} / h}$.}
    \label{fig:reynolds}
\end{minipage}%
\begin{minipage}[t]{.05\textwidth}
~
\end{minipage}%
\begin{minipage}[t]{.475\textwidth}
  \centering
    \hspace{-0.025\textwidth}
    \includegraphics[height=1.3\textwidth]{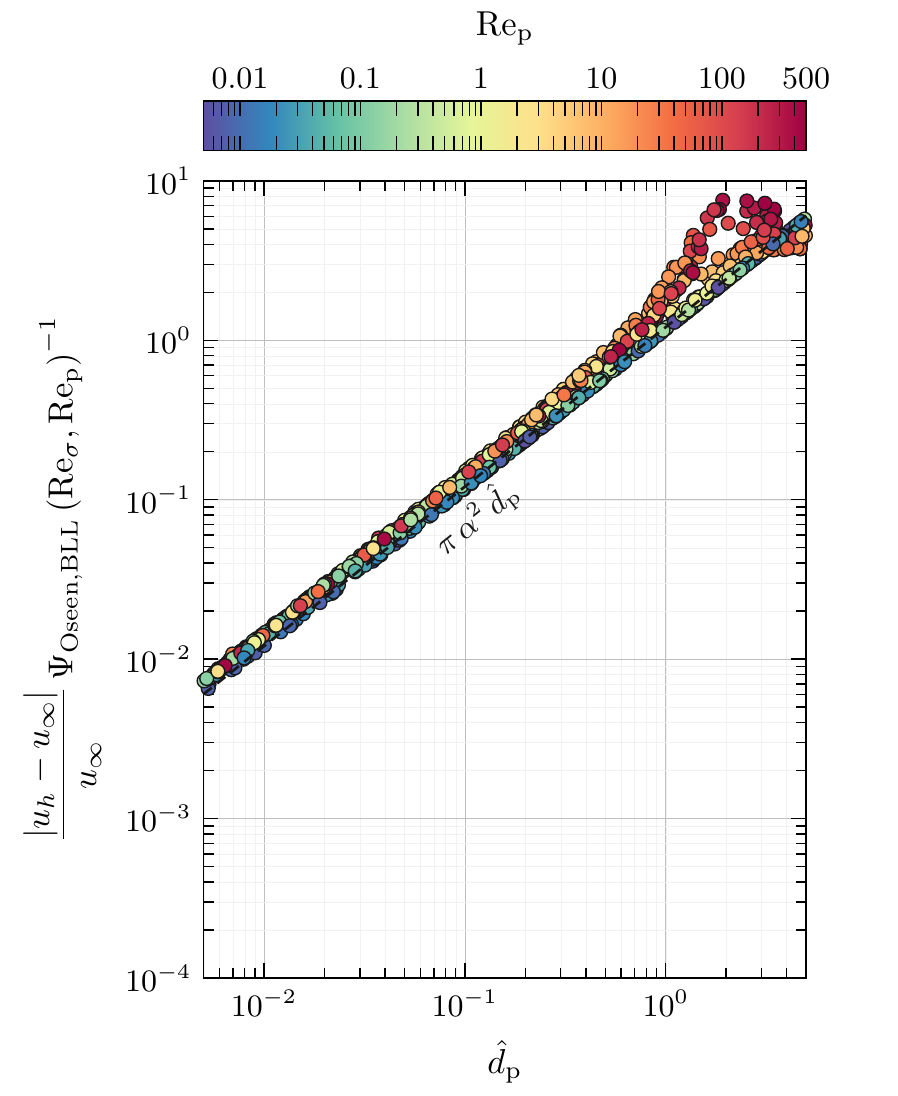}
    \captionof{figure}{Discrete velocity disturbances observed in the cell where the particle is located, using our numerical framework \textit{MultiFlow}, normalised by the far-field velocity $u_\infty$ multiplied by the Oseen correction function of \citet{Balachandar2019} , Eq.~\eqref{eq:oseenbll}, with the ratio $\smash{\hat{\sigma} = \sigma / h = 1/2}$.}
    \label{fig:bala}
\end{minipage}
\end{figure}

    Figure~\ref{fig:reynolds} shows the evolution of $\smash{\text{Re}_h}$ as a function of $\smash{\hat{d}_\text{p}}$ and $\smash{\text{Re}_\text{p}}$, as given by Eq.~\eqref{eq:reh}. We observe that $\smash{\text{Re}_h}$ scales proportionally to $\smash{\hat{d}_\text{p}}$ as this ratio decreases, which explains the good performance of Oseen's approximation even for large values of $\smash{\text{Re}_\text{p}}$, as long as $\smash{\hat{d}_\text{p}}$ remains small. When $\smash{\text{Re}_\text{p} = 500}$, for instance, $\smash{\text{Re}_h}$ is in fact smaller than $1$ for $\smash{\hat{d}_\text{p} < 0.01}$. In addition to these considerations, as pointed out by \citet{Balachandar2019}, the scaling based on Oseen's approximation can be expected to be accurate even for relatively high values of the Reynolds number, due to the relative simplicity and linearity of the flow induced by the momentum transfer between the Lagrangian particle and the Eulerian fluid.
\subsection{Alternative scaling based upon Gaussian regularisation} 
    \label{sec:discussionbala}
    Recently published works have addressed the estimation of a particle's self-induced flow disturbance in the frame of the \textit{volume-filtered} (or \textit{regularised}) EL method \cite{Gualtieri2015,Horwitz2016,Ireland2017,Horwitz2018,Battista2019,Balachandar2019,Poustis2019,Pakseresht2020,Evrard2020b}. In \cite{Balachandar2019}, particularly, \citeauthor{Balachandar2019} derive an expression for the local flow disturbance induced by the momentum fed back to the fluid by a Lagrangian particle, when the momentum source is convoluted with a Gaussian kernel. They show that the disturbance derived in the Stokes limit can be extended to finite Reynolds numbers by its scaling with the function
    \begin{equation}
        \Psi_\text{Oseen,BLL}(\text{Re}_\sigma,\text{Re}_\text{p}) = \dfrac{3\, f\!\left(\text{Re}_\text{p}\right)}{\text{Re}^3_\sigma\sqrt{2\pi}} \left( \pi - \text{Re}_\sigma\sqrt{2\pi} + \text{Re}_\sigma^2 \pi / 2 - \pi \exp\left(\text{Re}_\sigma^2 / 2\right) \text{erfc}\left(\text{Re}_\sigma / \sqrt{2}\right)\right) \, ,
        \label{eq:oseenbll}
    \end{equation}
    where $\sigma$ is the standard deviation of the Gaussian kernel. Applying this scaling to the object of the present study, that is the estimation of the velocity disturbance in the mesh cell containing a Lagrangian particle under the assumptions of the PSIC model, requires to relate the scale of the Gaussian kernel, $\sigma$, to that of the mesh cell, $h$. A value of the ratio $\hat{\sigma} = \sigma/h$ that allows to apply this scaling in the framework of the PSIC model is not provided by \citet{Balachandar2019}, so we propose to choose it as 
    \begin{equation}
        \arg\min_{\hat{\sigma}} \sum_{N_\text{s}} \left[\dfrac{|u_h - u_\infty|}{u_{\infty}} - \pi\,\alpha^2\,\hat{d}_\text{p}\,\Psi_\text{Oseen,BLL}\left(\hat{\sigma}\,\text{Re}_\text{p}/\hat{d}_\text{p}, \text{Re}_\text{p}\right) \right]^2 \ ,
    \end{equation}
    based on the results of the numerical experiments presented in Section~\ref{sec:results}. 
    For the $N_\text{s} = 800$ test-cases conducted in this study, we find the optimal ratio to be $\hat{\sigma} \simeq 1/2$. With this ratio, the discrete velocity disturbance normalised by the far-field velocity multiplied by $\smash{\Psi_\text{Oseen,BLL}(\hat{\sigma}\,\text{Re}_\text{p}/\hat{d}_\text{p}, \text{Re}_\text{p})}$, shown in Figure~\ref{fig:bala}, qualitatively collapses onto the dashed line $\smash{\propto \pi\,\alpha^2\,\hat{d}_\text{p}}$ (as previously observed in Figure~\ref{fig:errorsOS}).
    This alternative scaling, although also based on Oseen's approximation, differs from the one we propose in Section~\ref{sec:derivation} in several regards. Our approach, which yields Eq.~\eqref{eq:oseencorr}, considers the singular Oseenlet solution and integrates it in a sphere of volume $h^3$ to emulate the ingrained averaged nature of the solution available in each computational cell. The scaling proposed in~\cite{Balachandar2019}, on the other hand, considers the \textit{regularised} Oseenlet solution, which is non-singular at the center of the particle, but cannot be analytically integrated over a volume equivalent to that of the computational mesh cell. Since relying on the regularisation of the momentum transfer by its convolution with the Gaussian kernel, this latter scaling requires to relate the standard deviation of the Gaussian, $\sigma$, to the mesh-spacing, $h$, in order to be applicable within the PSIC-EL framework. As no direct and obvious way to relate these two quantities can be derived, an optimal value for the ratio $\hat{\sigma} = \sigma/h$ has to be found from the solution of a minimisation problem based on the results of numerical experiments.

\section{Summary and conclusions}
\label{sec:conclusions}
When tracking a particle within the PSIC-EL framework, the velocity disturbance induced by the momentum fed back to the flow by the particle is responsible for errors in the estimation of the forces acting on the particle. If drag is the dominant fluid force acting on the particle, the error in the estimation of the undisturbed flow velocity -- therefore the error in the estimation of drag -- is approximated as
\begin{equation}
    \text{Error} \simeq 
    \pi\,\alpha^2\,\hat{d}_\text{p}\,\Psi_\text{Oseen}\left( \text{Re}_\text{p}, \hat{d}_\text{p} \right) \, ,
\end{equation}
with
\begin{equation}
    \Psi_\text{Oseen}\left( \text{Re}_\text{p}, \hat{d}_\text{p} \right) = 3\,\hat{d}_\text{p}\,f\!\left(\text{Re}_\text{p}\right) \left( (\alpha\,\text{Re}_\text{p})^{-1} 
- 2\, \hat{d}_\text{p}\,(\alpha\,\text{Re}_\text{p})^{-2} + 2\,\hat{d}_\text{p}^{\,2}\,(\alpha\,\text{Re}_\text{p})^{-3} \left(1 - \exp\left( -\dfrac{\text{Re}_\text{p}\,\alpha}{\hat{d}_\text{p}} \right) \right) \right) \, ,
\end{equation}
and where $\smash{\alpha = \left(3/(4\,\pi)\right)^{1/3}}$, $\smash{\hat{d}_\text{p}}$ is the ratio between the particle diameter $\smash{{d}_\text{p}}$ and the mesh spacing $h$, $\smash{\text{Re}_\text{p}}$ is the particle Reynolds number, and $\smash{f\!\left(\text{Re}_\text{p}\right)}$ is an empirical factor extending Stokes' law to finite Reynolds numbers.
In the Stokes limit, this reduces to
\begin{equation}
    \text{Error} \simeq 
    (6/5)\,\hat{d}_\text{p} \, .
\end{equation}
Validated with the results of numerical simulations, these expressions have been shown to be accurate for $\smash{\hat{d}_\text{p} \lesssim 1/2}$ and $\smash{\text{Re}_\text{p} < 500}$.
As an example, for a particle that is 10 times smaller than the mesh spacing, which is commonly advised in the literature, drag will be estimated with an error of up to $\smash{12\,\%}$ in the Stokes limit. Using the correlation of \citet{Schiller1933} to extend Stokes' law, this error reduces to about $\smash{5\,\%}$ for $\smash{\text{Re}_\text{p} = 1}$, and to about $\smash{0.3\,\%}$ for $\smash{\text{Re}_\text{p} = 100}$.

\section{Acknowledgements}

We thank Dr. J\"org Schulenburg and Daniel Hasemann for their support with the computing systems at Otto-von-Guericke-Universit\"at Magdeburg.

{
}

\end{document}